\documentclass[fleqn,twoside]{article}
\usepackage{espcrc2}

\usepackage{graphicx}

\newcommand{\AmS}{{\protect\the\textfont2

  A\kern-.1667em\lower.5ex\hbox{M}\kern-.125emS}}

\title{Lattice study of ``f$_{0}$(600) or $\sigma$" }

\author{SCALAR Collabolation:\\
Teiji Kunihiro\address{YITP, Kyoto University, Kyoto 606-8502, Japan},
Shin Muroya\address{Tokuyama Women's College Tokuyama 745-8511, Japan}\thanks{Presenter},
Atsushi Nakamura\address[IMC]{IMC, Hiroshima
University, Higashi-Hiroshima 739-8521,  Japan },
Chiho Nonaka\address{Radiation Lab, RIKEN, 2-1 Hirosawa, Wako 351-0198, Japan}
\thanks{Present address: Department of Physics, Duke University, Durham, NC27708-0305, USA},
Motoo Sekiguchi\address{Faculty of Engineering, Kokushikan University,
Tokyo 154-8515,
Japan} \\
and 
Hiroaki Wada\address{Laboratory of Physics, College of Science and
Technology,
Nihon University, Chiba 274-8501, Japan}
        }

\begin{document}

\begin{abstract}
We investigate the propagator of  ``f$_{0}$(600) or the $\sigma$" by the
full-QCD simulation with Wilson fermions.  
We calculate the mesonic correlator in the $I=0$, $J^P=0^{+}$ channel on the $8^{3} \times 16$ lattice.  
Plaquet action and Wilson fermion action are adopted.
A coupling constant  $\beta$ is set to
 4.8 and three kinds of  hopping parameter, $\kappa$=0.1846, 0.1874 and 0.1891 are assayed.   
The disconnected diagram in the propagator is evaluated through taking
average over  500 or 1000 Z2 noise.
Simulations with the larger hopping parameter provide us 
with less noisy results.
Though the statistics is not yet enough, our results indicate the existence of a pole
with a mass in  almost the same order as that  of the $\rho$.

\end{abstract}

\maketitle

\section{Introduction}

In many text books, we can find the intuitive explanation of the spontaneous 
breaking of the the chiral symmetry based on the titled ``Mexican Hat'' 
potential, of which the angular and
 radial modes correspond  
to the $\pi$ and  $\sigma$ mesons, respectively.  
Also in the the Nambu Jona-Lasinio model, 
the  $\sigma$-degree of freedom   plays an important role
for the chiral symmetry breaking of QCD in the low-energy region. 
Despite its importance,  the experimental 
identification of the light scalar meson is a 
long standing puzzle and the $\sigma$ had disappeared from the list of 
Particle Data Group (PDG) for over 20 years.
But, the scenes around the $\sigma$ are changing recently;  
re-analyses of $\pi$-$\pi$ scattering phase shift 
strongly suggest the existence of the $I=0$ and $J^{PC}=0^{++}$ meson and 
significant  contributions of the $\sigma$ in the decay channels of heavy
particles such 
as D$\to \pi \pi \pi$
are reported \cite{KEK,decay}.
The $\sigma$ may reveal itself also in $\Upsilon(3S) \to \Upsilon \pi \pi$ 
channel\cite{Ishida}.
  In  1996 PDG, ``$f_0$(400-1200) or $\sigma$" appeared bellow $1$ GeV mass 
region.  In the 2002 edition, it appears as  ``$f_0$(600) or $\sigma$" and  
as the 105th branch of the D$^{+}$ decay,
  ``D $\to \sigma + \pi$ " is cited \cite{PDG}.
Now it is an important task of the lattice calculation 
to confirm the properties of the $\sigma$ meson based on QCD.

In quenched QCD,  DeTar and Kogut first measured the $\sigma$ meson 
screening mass in a lattice simulation \cite{DeTar}, 
Alford and Jaffe discussed 
the possible light scalar mesons as 
$\bar{q}^2q^2$ states \cite{Alford}, and  
masses and mixing of $q\bar{q}$ states
 and a glueball have been investigated by Lee and Weingarten
\cite{Lee}.  
However, as we have already reported in the previous proceedings
\cite{Lat01}, 
the connected  and disconnected diagrams
 give almost the same amount of the contributions,
indicating that the dynamical quark effect seems to be  essential in the calculation.

McNeile and Michael computed the  mixed iso-singlet scalar masses of 
$q\bar{q}$ and glueball states in two kinds of situation, i.e,  
with and without the dynamical quark 
effects \cite{McNeile}. 
We focus our attention  to explore  whether the $\sigma$ 
meson exists below $1$ GeV in this work.  
Based on the quite similar motivation, 
Prelovsek presented in the present conference 
the calculation of the quenched scalar meson
correlator 
with the domain wall fermions, 
where the disconnected diagram is taken into account 
effectively through an approximate formula based on the chiral 
perturbation theory \cite{Sasa}.

\vspace{-0.1cm}
\section{Formulation of the $\sigma$ propagator}

As for the $\sigma$ meson, we adopt the operator as
\begin{equation}
\sigma(x) \equiv
 \sum_{c=1}^3\sum_{\alpha=1}^4 
\frac{\bar{u}_\alpha^c(x)u_\alpha^c(x)+\bar{d}_\alpha^c(x)d_\alpha^c(x)}
{\sqrt{2}} \ ,
\end{equation}
where $u$ and $d$ are the $u$-quark and $d$-quark Dirac spinors, 
respectively. The indices $c$ and $\alpha$ denote color and Dirac spinor indices, 
respectively.   Quantum numbers of the operator are $I=0$ and $J^{P}=0^{+}$. 
Because our simulation is full QCD, 
any kind of the state which has the same quantum number 
can contribute to the correlation  as a virtual state.
 The $\sigma$ meson propagator is given by
\vspace{-0.2cm}
\begin{eqnarray}
&& G(y,x) = \langle \sigma(y) \sigma(x)^\dagger \rangle 
\ \ \ \ \ \ \ \ \  \ \
\ \ (y_0>x_0) \nonumber\\
&=& \frac{1}{Z} \int DUD\bar{u}DuD\bar{d}Dd \nonumber\\
&&\sum_{a,b=1}^3 \sum_{\alpha,\beta=1}^4 
\frac{\bar{u}_\beta^b(y)u_\beta^b(y)+\bar{d}_\beta^b(y)d_\beta^b(y)}
 {\sqrt{2}} \nonumber \\
&\times & \left( 
\frac{\bar{u}_\alpha^a(x)u_\alpha^a(x)+\bar{d}_\alpha^a(x)d_\alpha^a(x)}
{\sqrt{2}} 
\right)^\dagger 
e^{-S_g-\bar{u}Wu-\bar{d}Wd} \nonumber\\
&& \\
&=& - \langle {\mbox Tr} W^{-1}(x,y) W^{-1}(y,x) \rangle \nonumber \\
& & + 2 \langle {\mbox Tr} W^{-1}(y,y) {\mbox Tr} W^{-1}(x,x) \rangle \ .
\end{eqnarray}
In Eq.(3), "Tr" represents summation 
over color and Dirac spinor indices. 
$W^{-1}$'s are $u$ and $d$ quark propagators, $U$ is the link variable of
gluon, and $S_g$ is the pure gauge action of gluons. 
We assume that the $u$ and $d$ quark propagators are equivalent 
because $u$ and $d$ quark masses are almost the same. 
From Eq.(3), we can see that 
$\sigma$ propagator is composed 
of the connected diagram (the first term of Eq.(3)) and  the
disconnected one (the second term of Eq.(3)). 
The quantum number of the $\sigma$ meson ($I=0$ and $J^P=0^+$) 
is the same as that of the vacuum, and the vacuum expectation 
value of the $\sigma$ operator, $\langle \sigma(x) \rangle$,  does 
not vanish.  
Therefore, the contribution of $\langle \sigma(x) \rangle$ 
should be  
subtracted from the $\sigma$ operator.
Thus,
\begin{eqnarray}
G(y,x) &=& - \langle {\mbox Tr} W^{-1}(x,y) W^{-1}(y,x) \rangle \nonumber \\
& & + 2 \langle {\mbox Tr} W^{-1}(y,y) {\mbox Tr} W^{-1}(x,x) 
\rangle \nonumber \\
& & - 2 \langle {\mbox Tr} W^{-1}(y,y) \rangle \langle {\mbox Tr} 
W^{-1}(x,x) \rangle
\label{Eq-6}
\end{eqnarray}
is adopted for the $\sigma$ propagator. 
The third term in Eq.(4) corresponds to the subtraction of the vacuum
contribution. 
Our results show that  
the values of the second  and the third terms in Eq.(4) are 
in the same order.   Therefore, in order to obtain the signal  correctly 
as the difference between these terms, 
the high precision numerical simulations and careful analyses are 
required.

\vspace{-0.2cm}
\section{Results of numerical simulations}
We calculate the $\sigma$ propagator in the full QCD by using the 
Hybrid Monte Carlo (HMC) algorithm, 
in order to take the dynamical quarks which are important to 
estimate the disconnected diagrams of the $\sigma$ propagator. 
We use the $Z_2$ noise method to calculate the disconnected diagrams 
(the second and third terms in Eq.(\ref{Eq-6})). 
The 500 or 1000 random $Z_2$ numbers are generated.
The two-flavor Wilson fermion system 
is simulated on the $8^3 \times 16$ lattice. 

After 500 trajectories are updated in the quenched QCD, 
we start to update the configuration in the full QCD by using HMC algorithm.
More than 500 trajectories by HMC are spent for thermalization.  
The $\sigma$ propagators are calculated on a configuration in every 10 trajectories.

Based on ref.\cite{CPPACS}, we set 
$\beta = 4.8$ ($a=0.197(2)$ fm, $\kappa_c=0.19286(14)$). 
 Figures 1, 2 and 3 correspond to the  three values of
the hopping parameter, 
$\kappa=0.1846$(fig.\ 1), 0.1874 (fig.\ 2) and 0.1891(fig.\ 3), respectively. 
The numerical results are summarized in  Table 1.

\begin{figure}
\begin{center}
\includegraphics[width=.9 \linewidth]{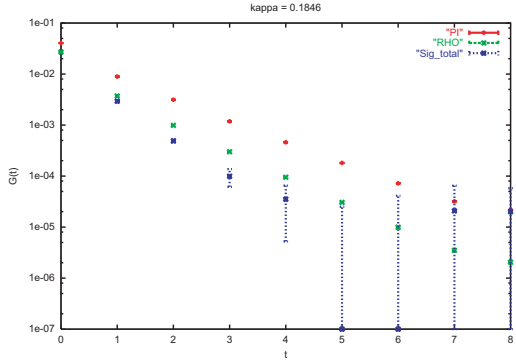}
\end{center}
\vspace{-15mm}
\caption{Propagators of $\pi$, $\rho$ and $\sigma$  for $\kappa=0.1846$
with periodic boundary condition.}
\label{k1846}
\end{figure}
\begin{figure}
\begin{center}
\includegraphics[width=.9 \linewidth]{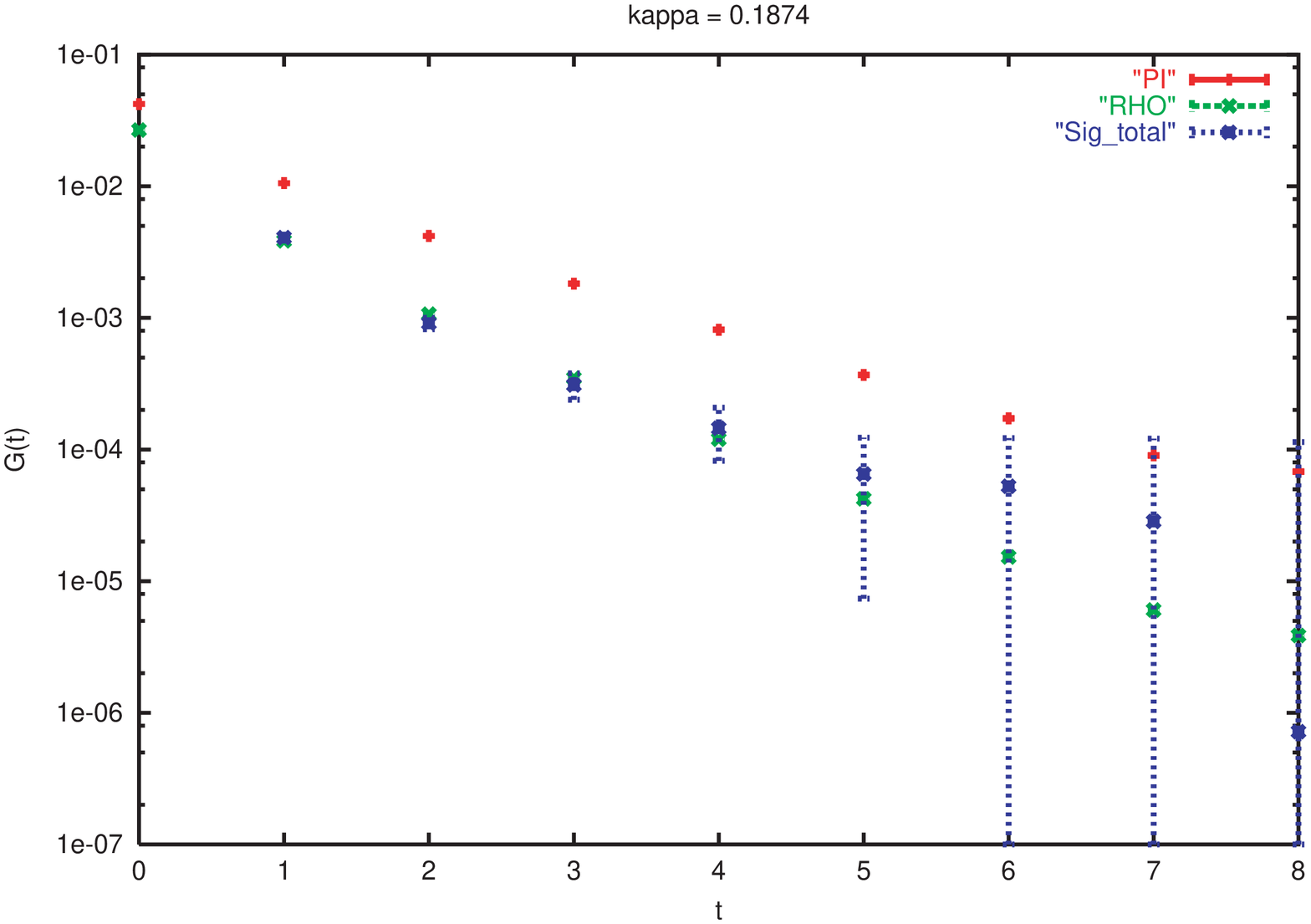}
\end{center}
\vspace{-15mm}
\caption{Propagators of $\pi$, $\rho$ and $\sigma$  for $\kappa=0.1874$
with periodic boundary condition.}
\label{k1846}
\end{figure}
\begin{figure}
\begin{center}
\includegraphics[width=.9 \linewidth]{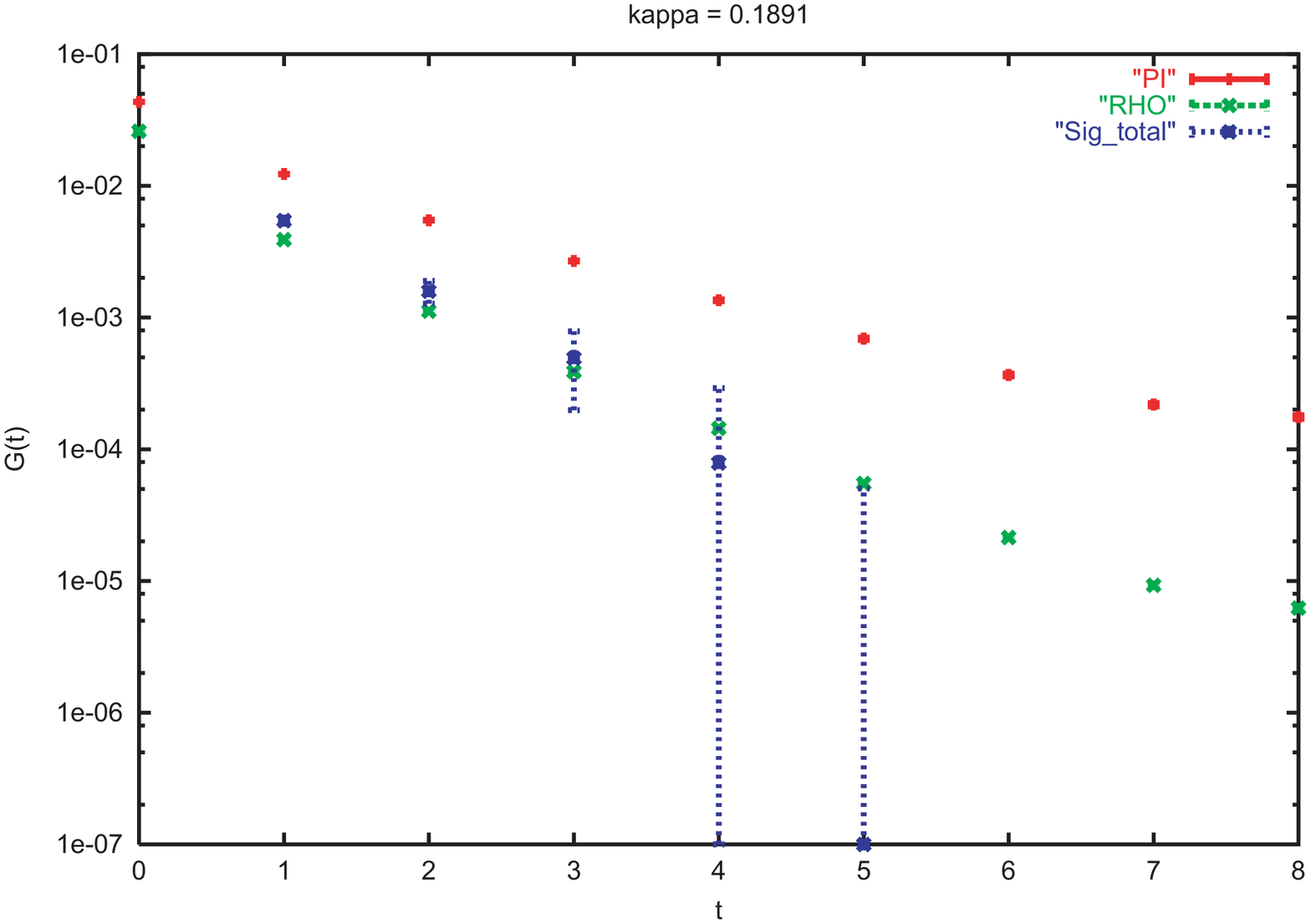}
\end{center}
\vspace{-15mm}
\caption{Propagators of $\pi$, $\rho$ and $\sigma$  for $\kappa=0.1891$
with periodic boundary condition.}
\label{k1846}
\end{figure}


\begin{table}
\caption{Summary of the results}
\begin{tabular}{c|c|c|c} 
\hline 
\hline 
$\kappa$& 0.1846 & 0.1874 &0.1891 \\ \hline
statistics $^{1)}$& 1380 & 900 &180  \\ \hline
$m(\pi)/m(\rho)$ $^{2)}$& 0.8291(12) &  0.7715(17) & 0.7026(32) \\ \hline
$m(\pi)/m(\rho)$ $^{3)}$ & 0.815&  0.747 &  0.68 \\ \hline
$m(\sigma)/m(\rho)$$^{3)}$  &1.2 & 0.96 & 1 \\ \hline
\end{tabular}
$^{1)}$number of configulations, $^{2)}$CPPACS, \\$^{3)}$our result
\end{table}
\vspace{-2mm}

\section{Concluding Remarks}

We have reported our preliminary results on 
the $\sigma$ meson propagator 
based on a full QCD lattice  calculation with dynamical fermions.
The simulations  with larger  hopping parameters give us clearer 
signals.   
Though our simulation is still far from the chiral limit and no
decay channel is yet open,
our preliminary results obtained with still a relatively low statics
indicate the existence of the light  
 $\sigma$ meson with the  mass in almost the same order as that 
of  the $\rho$ meson.  See Table 1.

\noindent
{\bf Acknowledgment}
This work is supported by Grant-in-Aide for Scientific Research by
Monbu-Kagaku-sho (No.\ 11440080, No.\ 12554008, No.\ 12640263 and
No.\ 14540263) and ERI of Tokuyama Univ.
Simulations were performed on SR8000 at IMC, Hiroshima
Univ., SX5 at RCNP, Osaka Univ., SR8000 at KEK.

\vspace{-4mm}

\end{document}